\documentstyle[12pt,fleqn]{article}
\renewcommand{\baselinestretch}{1.5}      
\def\title{\par\bigskip\begin{center}\bf\LARGE}
\def\endtitle{\end{center}\par\bigskip\par}
\def\instit{\begin{center}\it}
\def\endinstit{\end{center}}
\def\references{\begin{thebibliography}{10}}
\def\endreferences{\end{thebibliography}}
\renewcommand{\author}[1]{\begin{center}\Large #1\end{center}}
\renewcommand{\date}[1]{\par\bigskip\par\sl\hfill
#1\par\medskip\par}

\newcommand{\babs}{\hrule\par\begin{description}\item{Abstract:
}\it}
\newcommand{\eabs}{\par\end{description}\hrule\par\medskip}

\newcommand{\hs}{\hspace{2cm}}         
\newcommand{\nn}{\nonumber}            
\newcommand{\ap}{\left.}               
\newcommand{\at}{\left(}               
\newcommand{\ag}{\left\{}              
\newcommand{\cp}{\right.}              
\newcommand{\ct}{\right)}              
\newcommand{\cg}{\right\}}             
\newcommand{\beq}{\begin{equation}}                    
\newcommand{\eeq}{\end{equation}}                      
\newcommand{\beqn}{\begin{eqnarray}}                   
\newcommand{\eeqn}{\end{eqnarray}}                     
\newcommand{\ii}{\infty}                         
\newcommand{\fr}[2]{\mbox{$\frac{#1}{#2}$}}      
\renewcommand{\Re}{\,\mbox{Re}\,}                
\renewcommand{\Im}{\,\mbox{Im}\,}                

\newcommand{\be}{\beta}

\newcommand{\ze}{\zeta}

\newcommand{\Ga}{\Gamma}

\newcommand{\La}{\Lambda}

\newcommand{\Om}{\Omega}

\begin{document}


\begin{center}

{\LARGE \bf
Mellin-Barnes Representation for the Genus-$g$  Finite Temperature
String Theory }

\vspace{4mm}

\renewcommand
\baselinestretch{0.8}

{\sc A.A. Bytsenko} \\ {\it Department of Theoretical Physics,
State
Technical
University, \\ St Petersburg 195251, Russia} \\
{\sc E. Elizalde}\footnote{E-mail address: eli @ ebubecm1.bitnet}
\\
{\it Department E.C.M., Faculty of Physics, University of
Barcelona, \\
Diagonal 647, 08028 Barcelona, Spain} \\
{\sc S.D. Odintsov}\footnote{E-mail
address: odintsov @ theo.phys.sci.hiroshima-u.ac.jp}\\
{\it Dept. of Math., Pedagogical Institute, 634041 Tomsk,
Russia} \\
and {\it Department of Physics, Faculty of Science, Hiroshima
University, \\
Higashi-Hiroshima 724, Japan} \\
 {\sc S. Zerbini} \footnote{E-mail address: zerbini @ itncisca}\\
{\it Department of Physics, University of Trento, 38050 Povo,
Italy}    \\ and
{\it I.N.F.N., Gruppo Collegato di Trento}

\renewcommand
\baselinestretch{1.4}

\vspace{5mm}

\end{center}
 \begin{abstract}
 The Mellin-Barnes representation for the free energy of the
genus-$g$ string
is constructed. It is shown that the interactions of the open
bosonic
string do not modify the critical (Hagedorn) temperature. However,
for the
sectors having a spinor structure, the critical temperature exists
also for all $g$ and depends on the windings. The appearance of a
periodic structure is briefly discussed.
\end{abstract}

\vspace{8mm}


\vspace{1.5cm}

\newpage

The increasing interest in string theory at non zero temperature
[1-3] as well as
the numerous attemps to understand the origin and physical
interpretation of the Hagedorn temperature [2] (a typically stringy
phenomenon) have different motivations. If string
theory is to be considered as the main ingredient of the ``theory of
everything", then hot (super)strings should describe the dynamics
of the early universe. But even if strings were not so fundamental
as hoped by somebody, they certainly may be quite useful in connection
with QCD at non zero temperature.

The physical meaning of the Hagedorn temperature as the critical
one corresponding to the behavior of thermodynamic ensembles, may
be grasped by
investigating the interplay between free strings and their
interactions (i.e. higher
loops). In recent papers [4],  one-loop string
theory has been studied by making use of the Mellin-Barnes
representation for the related free energy.
Working with these techniques, a novel representation of the free
energy has been
obtained (Laurent series representation). In addition, the
Hagedorn temperature has been clearly interpreted as the
convergence radius of this Laurent series.

The aim of the present letter is to generalize the Mellin-Barnes
representation obtained in [4] to arbitrary, genus-$g$ strings.
Such a generalization will allow us to identify the critical
temperature
at arbitrary loop order. Early attemps to study the critical
temperature for multi-loop strings can be found in
[5-7].

 It is  well-known that the genus-$g$ temperature contribution to
the
free energy for the bosonic string can be written as [5]
\beq
F_{g}(\be) = {\sum^\ii_{m_i,n_j=-\ii}}' \int (d\tau)_{WP}\, (\det
P^+P)^{1/2}(\det
\Delta_g)^{-13}
 e^{-\Delta S(\be,\vec m,\vec n)}\,,
\label{1}
\eeq
where $(d\tau)_{WP}$ is the Weil-Petersson measure on the
Teichm\"{u}ller
space. This measure as well as the factors $\det (P^+P)$ and $\det
\Delta_g$ are each individually modular invariant [5].
In addition,
\beq
I_g(\tau)=(\det P^+P)^{1/2}(\det
\Delta_g)^{-13}=e^{c(2g-2)}Z'(1)^{-13}Z(2)\,,
\label{2}
\eeq
where $Z(s)$ is the Selberg zeta function and $c$  an absolute
constant [8]. Furthermore, the winding-number factor
has the form of a metric over the space of windings, namely
\beq
 \Delta S(\be,\vec m,\vec n)=\frac{T\be^2 }{2}[m_l\Omega_{li}-n_i]
((\Im \Om)^{-1})_{ij}
[\bar{\Omega}_{jk}m_k-n_j]=g^{\mu\nu}(\Om)N_{\mu}N_{\nu},
\label{3}
\eeq
$T$ being the string tension and $\mu,\nu=1,2,..,2g$,
$\{N_1,...,N_{2g}\}\equiv \{m_1,n_1,...,m_g,n_g\}$. The periodic
matrix $\Omega$, corresponding to the string world-sheet of genus
$g$,
is a holomorphic function of the moduli, $\Om_{ij}=\Om_{ji}$
and $\Im \Om >0$. The matrix $\Omega$ admits a decomposition into
real symmetric $g \times g$ matrices: $\Omega =\Omega_1+i\Omega_2$. As a
result
\beq
g(\Omega_1+i\Omega_2)=\at
\begin{array}{cc}
\Omega_1\Omega_2^{-1}\Omega_1+\Omega_2 & -\Omega_1\Omega_2^{-1}\\
-\Omega_2^{-1}\Omega_1 & \Omega_2^{-1}
\end{array}
\ct.
\eeq
Besides, $g(\Om)=\hat{\La}^tg(\La(\Om))\hat{\La}$ [8],
where $\La$ is an element of the symplectic modular group
$Sp(2g,Z)$ and the associated tranformation of the periodic matrix
reads $\Om \mapsto \Om'=\La(\Om)=(A\Om+B)(C\Om+D)^{-1}$. As a
consequence, the winding factor
\beq
\hs   \sum _{\vec m,\vec n}'
\exp{[-\Delta S(\be,\vec m,\vec n)]}\nn
\eeq
is also modular invariant.

It can be shown that the $2g$ summations present in the expression
for
$F_g(\be)$ can be replaced by a single summation together with a
change in the
region of integration from the fundamental domain to the analogue
of
the strip $S_{a_1}$ related to the cycle $a_1$, whose choice is
entirely arbitrary [5,6]. Then, one has
\beq
F_{g}(\be) = \sum^\ii_{r=1} \int (d\tau)_{WP}\, I_g(\tau)
\exp {[-\fr{T\be^2r^2}{2}(\Omega_{1i}
((\Im \Om)^{-1})_{ij} \bar{\Omega}_{j1})]}.
\label{5}
\eeq
Let us now consider a different representation for the genus-$g$
free
energy. We shall generalize the method introduced in ref. [4] for
the $g=1$
case. The idea is to make use of the Mellin tranform of the
exponential factor, i.e.
\beq
e^{-v}=\frac{1}{2\pi i}\int_{c-i\ii}^{c+i\ii}ds\,  \Ga(s)v^{-s} ,
\label{6}
\eeq
with $\Re v>0$ and $c>0$. Therefore one arrives at
\beqn
{\sum _{\vec m,\vec n=-\ii}^\ii} '
\exp{[-\Delta S(\be,\vec m,\vec n)]}&=&{\sum _{\vec m,\vec
n=-\ii}^\ii} '
  \frac{1}{2\pi i}\int_{c-i\ii}^{c+i\ii}ds\,  \Ga(s)(\Delta S(\be,\vec
m,\vec
n))^{-s}\nn \\
 &=&\frac{1}{2\pi i}\int_{c-i\ii}^{c+i\ii}ds\,  \Ga(s)(\fr{T\be^2}{2})^{-s}
G_g(s;\Om),
\eeqn
where
\beq
G_g(s;\Om)\equiv \sum_{\vec N \in Z^{2g}/\{0\}} (\vec N^t \Om \vec
N)^{-s}
\label{8}
\eeq
and
\beqn
&& \sum^\ii_{r=1} \exp {[-\fr{T\be^2r^2}{2}(\Omega_{1i}
((\Im \Om)^{-1})_{ij} \bar{\Omega}_{j1})]} \nn \\
&&=\frac{1}{2\pi i}
\int_{c-i\ii}^{c+i\ii}ds\,  \Ga(s)\ze(2s)
(\fr{T\be^2}{2})^{-s}[\Omega_{1i}
((\Im \Om)^{-1})_{ij} \bar{\Omega}_{j1})]^{-s}.
\eeqn
Finally, using the formulae (8) and (10) in eqs. (1) and (6),
respectively, one obtains
\beq
  F_g(\be)=\frac{1}{2\pi i}\int_{c-i\ii}^{c+i\ii}ds\,
\Ga(s)(\fr{T\be^2}{2})^{-s}
\ag\int d(\tau)_{WP}\, I_g(\tau) G_g(s;\Om)\cg _{(Reg)}
\eeq
and
\beq
  F_g(\be)=\frac{1}{2\pi i}\int_{c-i\ii}^{c+i\ii}ds
\Ga(s)\ze(2s)(\fr{T\be^2}{2})^{-s}
\ag \int(d\tau)_{WP}\, I_g(\tau) [\Omega_{1i}
((\Im \Om)^{-1})_{ij} \bar{\Omega}_{j1}]^{-s} \cg _{(Reg)}.
\eeq
These are the main formulae which will be used for the evaluation of
the genus-$g$
string contribution. In order to deal with such expressions
the integrals on a suitable variable in $(d\tau)_{WP}$ should be
understood as the regularized ones. In this way the order of
integration may be interchanged.

As an example, let us consider first the $g=1$ case. It is well
known that [6,8]
\beq
(d\tau)_{WP}=\frac{d\tau_1d\tau_2}{2\tau_2^2}
\eeq
and
\beq
I_1(\tau)=-\mbox{Vol} \ (R^{26})(2\pi)^{-13}[\tau_2|\eta(\tau)|^4]^{-12}
\,, \eeq
where $\eta(\tau)=\exp(i\pi \tau/12)\prod_{n=1}^\ii[1-\exp(2 \pi i
n\tau)]$
is Dedekind's eta function.
In the case of an open bosonic string, we have $\Om_1=0$,
$\Om_2=\tau_2$,
and $\Om=$ diag $(\tau_2,\tau_2^{-1})$. In the limit $\tau_2
\rightarrow
0$ we  get
\beq
\exp {[-\fr{T\be^2} {2} (\vec N^t \Om \vec N)^{-s}]} \mapsto
\exp{(-\fr{T\be^2} {2}n^2\tau_2^{-1})}
\label{14}
\eeq
and
\beq
G_1(s;\Om) \mapsto
\sum^\ii_{n=1}(n^2\tau_2^{-1}) ^{-s}=\tau_2^s\zeta(2s) .
\label{15}
\eeq
The corresponding contribution to the free energy is given by
\beq
  \frac{1}{2\pi i}\int_{c-i\ii}^{c+i\ii}ds\,
\Ga(s)(\fr{T\be^2}{2})^{-s}\zeta(2s)
\ag \int_0^\ii d\tau_2 \tau_2^{s-14} \eta(i \tau_2)^{-24} \cg
_{(Reg)}.
\eeq
After having regularized the ultra-violet region ($\tau_2
\rightarrow 0$), one has (for more details see [4])

\beq
  \frac{1}{2\pi i}\int_{c-i\ii}^{c+i\ii}ds\,
\Ga(s)(\fr{T\be^2}{8\pi})^{-s}\zeta(2s) \Re (-1)^s \Ga(1-s).
\eeq

As a result, the one-loop free energy admits the representation
in terms of Laurent series :
\beq
F_1(\be)\sim \sum_{k=1}^\ii
\zeta(2k)y^{2k}+C_0(\mu)+C_1(\mu)(\be)^{-1}+F_R(\be,\mu) \, ,
\eeq
where
\beq
\hs y=(\fr{8\pi}{T})^{1/2}(\be)^{-1} \, ,
\label{}
\eeq
and the functions $C_0(\mu)$ and
$C_1(\mu)$ depend on the infrared cutoff parameter $\mu$, while $
F_R(\be,\mu)$ is the contribution coming from the contour integral
along the arc of radius $R$ on the right half-plane. In this
representation the critical (Hagedorn) temperature arises as the
convergence condition of the Laurent series, namely
\beq
\hs  \lim_{k \rightarrow \ii} \frac{\zeta(2k+2)}{\zeta(2k)}y^2=
y^2 < 1.
\label{18}
\eeq
As a consequence, $\be >\be_c=(\fr{8\pi}{T})^{1/2}$, $\be_c$ being the
inverse of the Hagedorn
temperature for the open (closed) bosonic string. Moreover, the
Laurent series can be summed and a periodic structure appears,
i.e.
\beq
F_1(\be)\sim y\cot (\pi y)
+C_0(\mu)+C_1(\mu)(\be)^{-1}+F_R(\be,\mu).
\label{19}
\eeq
This provides also a possible analytic continuation beyond the
Hagedorn temperature.

By analogy with the above one-loop evaluation, we shall consider in
the
following the open string genus-$g$ contribution to the free
energy.
The matrix $\Om$ may be chosen as  $\Om=diag(\Om_2,\Om_2^{-1})$. In
the limit $\Om_2 \rightarrow 0$, one has
\beq
\exp {[-\fr{T\be^2} {2} (\vec N^t \Om \vec N)^{-s}]} \mapsto
\exp{(-\fr{T \be^2}{2}\Om_2^{-1}\vec N^t \vec N)}\, ,
\label{20}
\eeq
and
\beq
G_g(s;\Om) \mapsto \Om_2^s
\sum_{\vec N \in Z^g/\{0\}}(\vec N^t \vec N) ^{-s}=\Om_2^s
Z_g|^{\vec 0}_{\vec 0}|(2s) \, ,
\label{21}
\eeq
where the Epstein zeta function of order $g$ is defined by
\beq
Z_g|^{\vec b}_{\vec h}|(s)=\sum_{\vec N \in Z^g/\{0\}}
[(n_1+b_1)^2+...+(n_g+b_g)^2]^{-s/2} \exp {[2\pi i (\vec N^t,\vec
h)]}\, .
\label{22}
\eeq
The corresponding contribution is given by
\beq
  \frac{1}{2\pi i}\int_{c-i\ii}^{c+i\ii}ds\,  \Ga(s)(\fr{T\be^2}{2})^{-s}
Z_g|^{\vec 0}_{\vec 0}|(2s)
\ag \int d\tau_{WP} \Om_2^{s}I_g(\tau)  \cg _{(Reg)}.
\eeq

Since a tachyon is present in the spectrum, the total free energy
will be divergent, for any $g$. The infrared divergence may be
regularized by means of a suitable cutoff parameter. This divergent
behavior
can be associated with the procedure of pinching a cycle non
homologous at zero (see for example [7]). It is well
known that the behavior of the factor $(d\tau)_{WP}\, I_g(\tau)$ is
given
by the Belavin-Knizhnik double-pole result and has a
universal character, for any $g$. It should also be noticed that
this
divergence is $\be$-independent and the meromormphic structure is
similar to genus-one case. As a consequence, the complete
expression for the free energy may be obtained again in terms of
the Laurent series, and the whole genus dependence of the critical
temperature is encoded in the Epstein zeta function
$ Z_g|^{\vec 0}_{\vec 0}|(2s)$.

For this reason, we have to determine the asymptotic properties of
$Z_g|^{\vec 0}_{\vec 0}|(2s)$. With this aim, we make use of the
following general result:
\beq
C_g\equiv \lim _{\Re s \rightarrow +\ii}
\frac{Z_g|^{\vec b}_{\vec 0}|(2s+2)}{Z_g|^{\vec b}_{\vec 0}|(2s)}=
[(\hat{b}_1-\eta_1)^2+...+(\hat{b}_g-\eta_g)^2]^{-1 } ,
\label{24}
\eeq
where at least one of the $b_i$ is noninteger,
$\hat{b}_i=b_i-[b_i]$
with $[b_i] $ the noninteger (decimal) part of $b_i$ and
\beq
\eta_i=\ag
\begin{array}{cc}
0 \, , &\hspace{1cm} 0\leq \hat{b}_i \leq 1/2, \\
1 \, , & \hspace{1cm} 1/2\leq \hat{b}_i < 1  .
\end{array} \right.
\eeq
Furthermore, if $\vec b=(0,0,...0)$, then $C_g=1$.
As a consequence, we arrive at the conclusion that the
interactions of bosonic strings do {\it not} modify the critical
Hagedorn temperature. This result is in agreement with other
computations [5,7].

However one can consider also different linear real bundles over
compact Riemann surfaces and spinorial structures on them. The
procedure of evaluation of the free energy in terms of the path
integral
over the metrics $g_{\mu\nu}$ does not depend on whatever type of
real
scalars are considered. This fact leads to new contributions to
the genus-$g$
integrals (1) and (2). On the other hand, one can investigate the
role
of these contributions for the torus compactification
[8,9]. In this case, the sum in eq. (1)
should be taken over the vectors on the lattice on which some space
dimensions are compactified. The half-lattice vectors can be
labelled
by the multiplets $(b_1,..,b_p)$, with $b_i=1/2$. The critical
temperature
related to the
multiplet $\vec b=(b_1,...b_p,0,..,0)$ can be easily evaluated by
means of eq. (27), which gives $C_p=4p^{-1}$. As a result
\beq
\hs   \hs   T_{c,p}=\frac{\sqrt p}{2} \, T_c.
\label{26}
\eeq
We note that  the particular multiplets
$(0,..,1/2,..,0)$ and $(0,..1/2,..,1/2,..0)$, where only one $b_i$
and two $b_i$ are different from
zero,
are associated with  ``minimal" critical temperatures given by $
T_{c,1}=T_c/2$ and  $T_{c,2}=T_c/{\sqrt 2}$ respectively.

Finally, making use of the following series representation for the
Epstein zeta function [10]
\beqn
Z_g|^{\vec 0}_{\vec 0}|(2s)&=& 2\frac{\pi^{\fr{g-1}{2}}}{\Ga(s)}
\sum_{k=1}^{g-1}\ag
\pi^{\fr{1-k}{2}}\Ga(s+\fr{k-g}{2})\zeta(2s+k-g)+\cp \\
 &+& \ap 2^{\fr{g+1-2s}{2}}(2\pi)^{\fr{(2-k)}{2}}\sum_{n=1}^\ii
\sum_{\vec N \in Z^{g-k}/\{0\}}(\fr{\vec N^t \vec N }{n^2}
)^{\fr{2s+k-g}{4}}
K_{\fr{g-2s-k}{2}}\left( 2\pi n \sqrt{\vec N^t \vec N}\right) \cg  , \nn
\eeqn
where $K_\nu(x)$ is the modified Bessel function, we conclude
that
it is possible to rewrite the integrand of eq. (26) in terms of the
finite sum appearing in eq. (30). As a consequence, a finite
temperature
periodic structure is also  present in the genus-$g$
contribution.

\vspace{5mm}

{\large \bf Acknowledgments}.
We would like to thank G. Cognola, L. Vanzo and K. Kirsten for
discussions. A.A.B. is grateful to I.N.F.N. gruppo collegato di
Trento for financial support.
E.E. has been supported by DGICYT (Spain), research project
PB90-0022, and by the Generalitat de Catalunya.
S.D.O. wishes to thank the Japan Society for the Promotion of
Science (JSPS, Japan) for financial support  and the
Particle Physics Group at Hiroshima University for kind
hospitality.


\end{document}